# Data-driven design of multilayer hyperbolic metamaterials for near-field thermal radiative modulator with high modulation contrast

Tuwei Liao[1], C. Y. Zhao[1], Hong Wang[2], and Shenghong Ju[1, 2, *]

[1] China-UK Low Carbon College, Shanghai Jiao Tong University, Shanghai, 201306, China

[2] Materials Genome Initiative Center, School of Material Science and Engineering, Shanghai Jiao Tong University, Shanghai 200240, China

**Abstract** The thermal modulator based on the near-field radiative heat transfer has wide applications in thermoelectric diodes, thermoelectric transistors, and thermal storage. However, the design of optimal near-field thermal radiation structure is a complex and challenging problem due to the tremendous number of degrees of freedom. In this work, we have proposed a data-driven machine learning workflow to efficiently design multilayer hyperbolic metamaterials composed of α-MoO$_3$ for near-field thermal radiative modulator with high modulation contrast. By combining the multilayer perceptron and Bayesian optimization, the rotation angle, layer thickness and gap distance of the multilayer metamaterials are optimized to achieve a maximum thermal modulation contrast ratio of 6.29. This represents a 97% improvement compared to previous single layer structure. The large thermal modulation contrast is mainly attributed to the alignment and misalignment of hyperbolic plasmon polaritons and hyperbolic surface phonon polaritons of each layer controlled by the rotation. The results provide a promising way for accelerating the designing and manipulating of near-field radiative heat transfer by anisotropic hyperbolic materials through the data-driven style.

---

[*] Corresponding author: shenghong.ju@sjtu.edu.cn

# 1. Introduction

Thermal radiation has been a fundamental way of heat transfer which is widely used in energy harvesting and management [1-6]. When the distance between the heat source and the heat sink is comparable to or smaller than the characteristic thermal wavelength, evanescent waves which provide photon tunneling will dominate the radiative heat transfer. The near-field radiative heat transfer (NFRHT) can exceed the blackbody limit by several orders of magnitude [7-9]. During the past two decades, the NFRHT has been studied and investigated because of its promising applications such as thermal diodes [10-12], thermal transistors [13-15], thermal photovoltaic [16-20], thermal switches [21, 22]. Great effort has been devoted to exploring various materials [23] and structures [24-26] to obtain large heat flux in NFRHT.

Recently, the hyperbolic materials (HMs) have attracted extensive research interest due to their hyperbolic dispersion characteristics, which arise from the components of their dielectric tensor with opposite signs of permittivity [27-29]. Due to their ability to support high wavevector modes, HMs exhibit significantly enhanced local optical density which effectively enhances the thermal radiation. Recent studies on the far-field thermal radiation have demonstrated that the layered structures composed of HMs possess remarkable physical properties [30, 31]. For the near-field thermal radiation, the hyperbolic bulk materials and stacked structures have been mainly investigated and discussed [30]. By exciting hyperbolic surface phonon polaritons (HSPhPs), hyperbolic phonon polaritons (HPPs) and localized polaritons (LPs), the heat flux in NFRHT can be significantly enhanced.

The near-field thermal radiation using HMs has been applied in thermal modulation in recent years [32-35]. At early stage, the modulation of radiative heat flux can be achieved by varying the gap distance between two materials. However, the precise control of the gap distance is strictly required to achieve the desired modulation target. Other efforts have also been devoted to improving the NFRHT modulation performance, such as manipulating the geometric structure and employing HMs as coatings. Biehs et al. have proposed a method to modulate radiative heat flux by tuning the relative orientation of two polarizing/metallic gratings [24]. Hu et al. proposed a method to achieve modulation of NFRHT using rotationally anisotropic HMs particles [33]. This alternative modulation technique can partially alleviate the drawbacks of modulators based on varying gap distances. However, the manufacture of grating structures or nano particles will be a significant challenge, as the roughness and manufacturing precision need to be ensured within several nanometers. Liu et al designed a NFRHT modulator based on hexagonal Boron Nitride (hBN), and the modulation contrast can reach higher than 5 for hBN films separated by a nanoscale gap distance [32]. With the rapid advancement of machine learning techniques, researchers have been exploring ML-assisted methods to accelerate the design of radiative structures in thermal radiation [36-39]. For instance, García-Esteban et al. [40] have demonstrated the effectiveness of simple neural network architectures trained with moderate-sized datasets, serving as fast and accurate surrogates for numerical simulations in radiative heat transfer. However, the application depends on the training data set, which is one of the main limitations and challenges for using neural networks to solve thermal-radiation problems. Wen et al. [41] have proposed a machine learning strategy that combines an artificial neural network with a genetic algorithm. Despite its efficiency, this strategy still presents challenges in terms of prediction errors between the predicted and accurate heat fluxes.

In this work, we have proposed a data-driven method to design the multilayer radiative modulator composed of hyperbolic material α-$MoO_3$. This contactless thermal modulator based on hyperbolic materials avoid manufacturing devices with complicated nanostructures. With the support of machine learning, the designed optimal multilayer metamaterials can reach high modulation contrast over 6.0 by smartly tuning the rotation, and the NFRHT heat flux is also larger compared with the hBN thermal modulators [32]. To reveal the physical mechanisms for the high modulation contrast, the effect of rotation angle, gap distance, and layer thickness on NFRHT are systematically investigated and discussed, and the energy transmission coefficient is also analyzed. Our work provides useful insight for accelerating the design of multilayer radiative modulator via the hybrid machine learning techniques.

**2. Computational details and machine learning algorithms**

For the thermal modulator based on multilayer metamaterials, due to the increase of the layer number, the design degrees of freedom of the geometric structures will significantly increase, making the optimal design more challenging. In this work, the non-contact near-field radiation thermal modulator based on multilayer α-MoO$_3$ metamaterials has been designed and investigated as shown in Fig. 1, where the emitter and receiver are separated by a vacuum gap $d$, and they are both composed of three layers and assumed to be semi-infinite. The emitter temperature $T_h$ and receiver temperature $T_l$ are set as 300 K and 0 K, respectively. To realize the efficient design of the proposed multilayer thermal modulator, serious number of structure parameters are considered here, including the gap distance $d$, the layer film thickness $t_i$ ($i$=1,2,3,4,5,6), and the rotation angles $\varphi_i$ ($i$=1,2,3,4,5,6). The detail tuning range of these parameters during the optimization process is listed in Table. 1. In this work, the optimization process considers intermediate integer angles from 0 to 90 degrees during the optimization process for each layer to explore various random angle combinations. The design target of the multilayer radiative modulator is to achieve the maximum modulation contrast controlled by the lattice rotation in each layer.

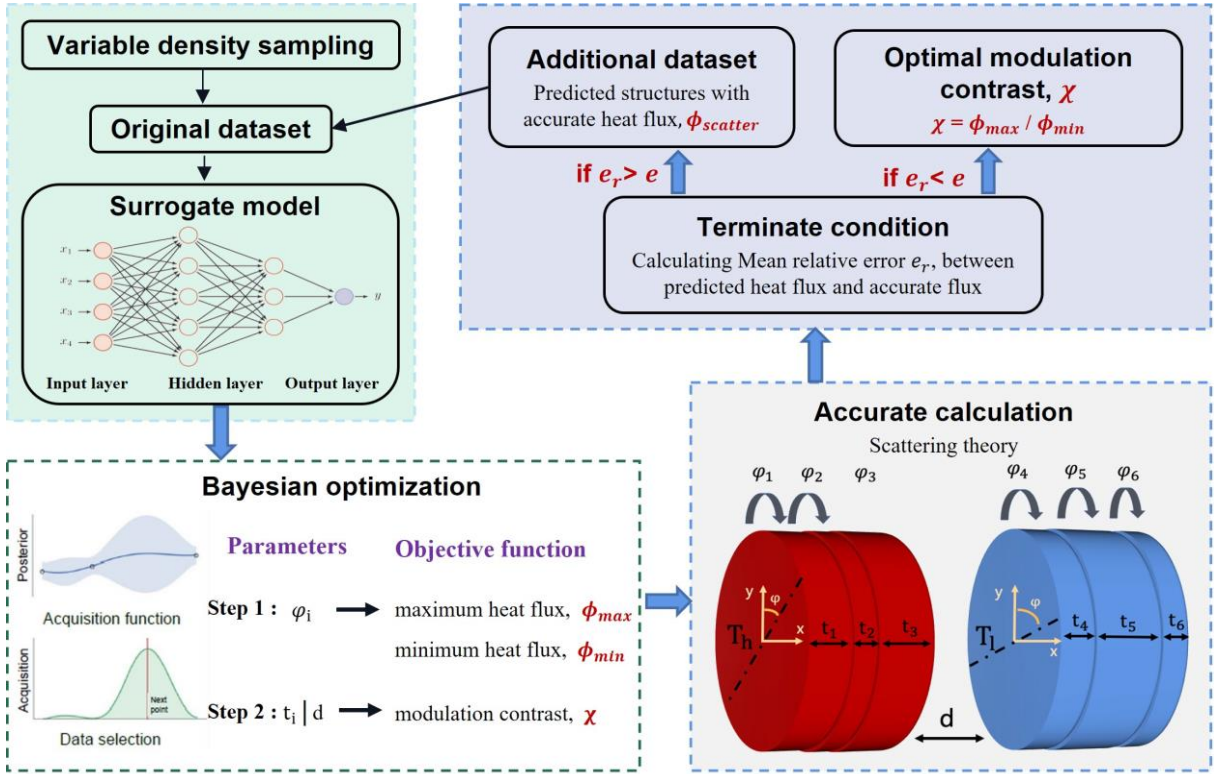

**Fig. 1.** Schematics of the data-driven design of multilayer hyperbolic metamaterials for near-field thermal radiative modulator. The target is to obtain the largest modulation contrast between multilayer hyperbolic metamaterials.

**Table. 1.** Parameters and optimization range for the design of multilayer hyperbolic metamaterials.

| Process | Range | Rotation angle $\varphi_i$ | Thickness $t_i$ / nm | Distance $d$ / nm |
|---------|-------|---------------------------|----------------------|-------------------|
| Step1   | Lower bound | 0 | (7) | (20) |
|         | Upper bound | 90 | (7) | (20) |
| Step2   | Lower bound |    | 1 | 1 |
|         | Upper bound |    | 100 | 100 |

The proposed data-driven design workflow for the multilayer metamaterials is by combining the multilayer perceptron (MLP) and Bayesian optimization (BO), as shown in Fig. 1. Firstly, the Latin hypercube sampling (LHS) method is adopted to select reasonable sampling points through the variable-density sampling which increases the sampling density at the high gradient regions of the model. These radiation properties of selected sampling points are then calculated by using accurate NFRHT modeling and computational methods, and are collect as initial training dataset. Secondly, the MLP neural network is trained as the surrogate functional model to connect the input feature vectors to corresponding output property. A two-step evaluation of the thermal radiative modulator is conducted here. For the step 1 of the optimization process shown in Fig. 1, the input parameters are the rotation angles $\varphi_i$ ($i$=1,2,3,4,5,6) of each layer, and the output is the NFRHT heat flux between the emitter and receiver. For the initial design stage, the gap distance is fixed at 20 nm and the thickness of each layer is fixed as 7 nm. For step 2, the input parameters are the thickness $t_i$ ($i$=1,2,3,4,5,6) of each layer and the gap distance between emitter and receiver, the output is the modulation contrast $\chi$ accordingly. Finally, the BO is combined with the MLP surrogate model and the Gaussian process is used to suggest a group of potential optimal structures. After calculating these suggested structures accurately, the new obtained data is added into the training dataset for updating the MLP surrogate model. The above-mentioned processes are repeated and iterated until the modulation contrast converges.

As an anisotropic material, the optical response of α-MoO$_3$ is dependent on the orientation of its principal axis. The principle components of α-MoO$_3$ relative dielectric constant can be described using the following Lorentz equations [42],

$$\varepsilon_m = \varepsilon_{\infty,m}\left(1 + \frac{\omega_{LO,m}^2 - \omega_{TO,m}^2}{\omega_{TO,m}^2 - \omega^2 - j\omega\Gamma_m}\right), \quad m = x, y, z \tag{1}$$

where $\omega$ is the angular frequency, and $x$, $y$, and $z$ represent the three principal axes corresponding to the crystalline directions [100], [001], and [010] of α-MoO$_3$ with lattice constants of $a$ = 0.396 nm, $b$ = 1.385 nm, and $c$ = 0.369 nm, respectively [43]. The detailed parameters used in our calculations are from Ref [42]. The real parts of the relative dielectric constant components are shown in Fig. 2(a). It can be seen that $\varepsilon_x$, $\varepsilon_y$ and $\varepsilon_z$ all shows negative values in the three spectral regions between 1.0273-1.6041×10$^{13}$ rad/s, 1.5457-1.8322×10$^{13}$ rad/s, and 1.8058-1.8925×10$^{13}$ rad/s, respectively. These three special regions correspond to the three Reststrahlen bands of α-MoO$_3$, in which the dispersion of electromagnetic waves can exhibit birefringent characteristics. The existence of the Reststrahlen bands significantly affect the non-contact heat transfer between α-MoO$_3$ bi-axial crystals, which will be discussed later.

The NFRHT along different crystal direction is considered in this work. When the crystalline direction [010] is perpendicular to the gap surface, the heat flux is assumed to be along the [010] direction. When tuning the rotation angle of each layer, two principal axes perpendicular to the heat flux direction are rotated by an angle $\varphi_i$ ($i$=1,2,3,4,5,6). Assuming that the relative dielectric constant tensor is $\varepsilon$, the relative dielectric constant tensor of each layer after rotation is given by [32],

$$\varepsilon_i = \begin{pmatrix} \cos\varphi_i & -\sin\varphi_i & 0 \\ \sin\varphi_i & \cos\varphi_i & 0 \\ 0 & 0 & 1 \end{pmatrix} \varepsilon \begin{pmatrix} \cos\varphi_i & \sin\varphi_i & 0 \\ -\sin\varphi_i & \cos\varphi_i & 0 \\ 0 & 0 & 1 \end{pmatrix} \tag{2}$$

Following the fluctuation-dissipation theorem and the reciprocity of the Green's function, the NFRHT heat flux between media 1 and 2 can be expressed as,

$$Q = \frac{1}{8\pi^3} \int_0^\infty [\Theta(\omega, T_1) - \Theta(\omega, T_2)]d\omega \int_0^{2\pi} \int_0^\infty \xi(\omega, \beta, \phi)\beta d\beta d\phi \tag{3}$$

where $\Theta(\omega, T) = \hbar\omega/(e^{\hbar\omega/k_BT} - 1)$ is the average energy of a Planck oscillator at temperature $T$, $\hbar$ is the normalized Planck's constant, $k_B$ is the Boltzmann's constant, $\beta$ is the parallel wave vector component, and $\phi$ is the azimuthal angle, $\xi(\omega, \beta, \phi)$ is referred as the energy transmission coefficient.

## 3. Results and discussion

We first consider the radiative modulator using single-layer of α-MoO$_3$. Figure 2 (b) and (c) show the modulation contrast and radiative heat flux versus the layer thickness and the gap distance. When the gap distance is fixed as 20 nm, the modulation contrast gradually increases to a maximum modulation contrast of 3.2 as the film thickness increases from 1 nm to 25 nm. As the layer thickness continues to increase, the control ratio begins to decrease. When the film thickness is thicker than 1000 nm, the modulation contrast convergences to a stable value of 1.9. Similarly, when the film thickness is fixed as 1000 nm, the thermal radiation heat flux significantly decreases as the gap distance increases, which indicates that gap distance has a significant impact on NFRHT. With the knowledge gained from the simple case, we further conduct the design the multilayer radiative modulator, which can further improve the modulation performance controlled via lattice rotation.

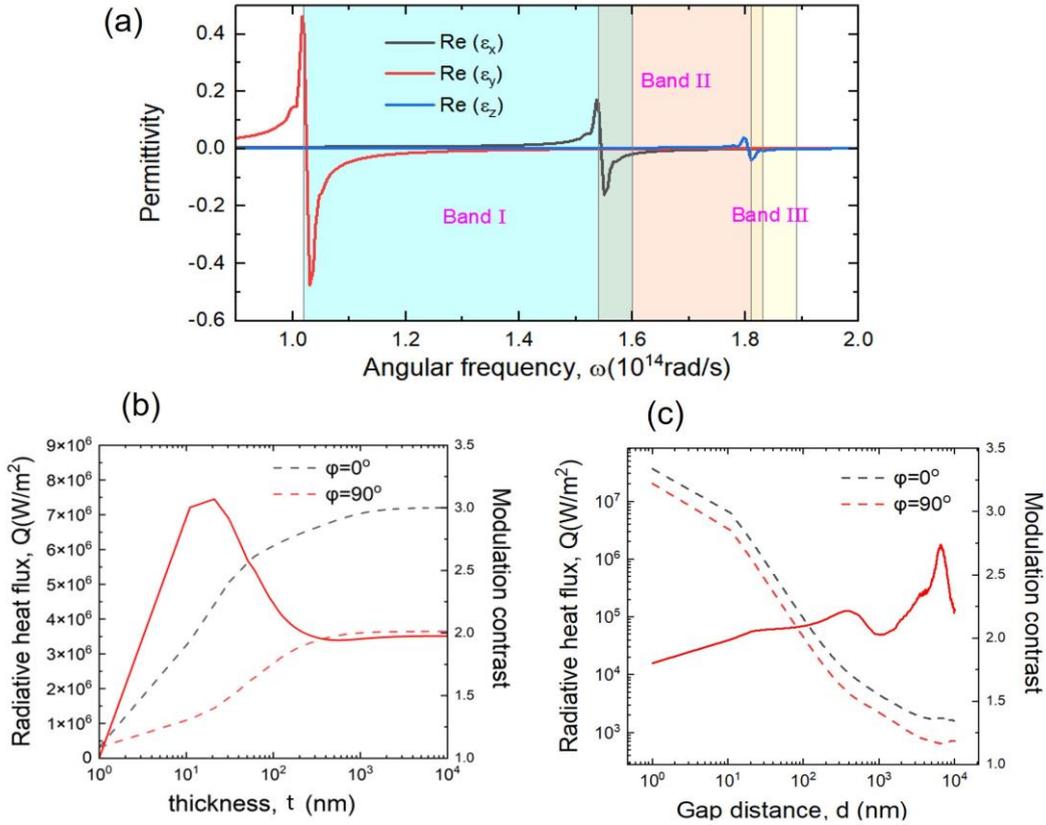

**Fig. 2.** (a) Real parts of the α-MoO$_3$ relative permittivity components varying with angular frequency. (b)-(c) show the modulation contrast and radiative heat flux between emitter and receiver with only single layer of α-MoO$_3$, (b) with varying layer thickness, and the gap distance is set as 20 nm, (c) with varying gap distance, and the thickness is fixed as 25 nm.

Figure 3(a)-(b) show the convergence of the modulation contrast versus the number of calculated structures by using the proposed hybrid MLP and BO design workflow. To demonstrate efficiency of the design framework, we also conduct optimization using pure BO and the results are shown in Fig.3 (b) for comparison. Three independent MLP+BO and BO optimization experiments with different random seeds are conducted respectively under the same parameter settings for fair comparison. The result shows that the modulation contrast for all rounds of MLP+BO optimizations eventually convergence to 6.29. On the contrary, the maximum modulation contrast by pure BO optimization is much smaller than that of MLP+BO within the same 3200 calculated structures. Figure 3(c)-(d) show the convergence of rotation angle of each layer varying with the number of calculated structures. For layers close to the near-field vacuum (layer 3 and 4), the angle reaches convergence with a faster speed. Conversely, layers far away from the vacuum (layer 1 and 6) exhibit a slower convergence speed.

To further illustrate the efficiency of proposed design framework, Table 2 compares the time consumption during the multilayer metamaterials design. By using a workstation with 2*Intel Xeon Gold 6271, the calculations of initial 2000 structures take about 33 hours. During each iteration with ten suggest structures, it takes approximately 10 minutes, including the consumed time for neural network model training by using the Levenberg-Marquardt (L-M) algorithm. Compared to the Bayesian regularization algorithm, the L-M algorithm has a faster training speed, taking about 1 minute per iteration. After establishing the surrogate model, the Bayesian optimization can figure out the optimal design parameters for thermal modulation within 30 seconds.

We next begin to design the radiation modulator by tuning purely the rotation angle $\varphi_i$ of each α-MoO$_3$ layer within the range of 0° to 90°. The gap distance $d$ is fixed at 20 nm and the thickness $t_i$ of each layer is fixed at 7 nm. The emitter and receiver temperature are set as 300 K and 0 K, respectively. The maximum and minimum heat flux obtained following the proposed optimization framework are 1882 kW/m$^2$ and 382 kW/m$^2$, respectively, with a rotation modulation contrast of 4.93. The detail optimal parameters after optimization are listed in Table. 3. Compared with the optimal single layer result shown in Fig. 2, the thermal modulation contrast of multilayer structure is further improved by 54%.

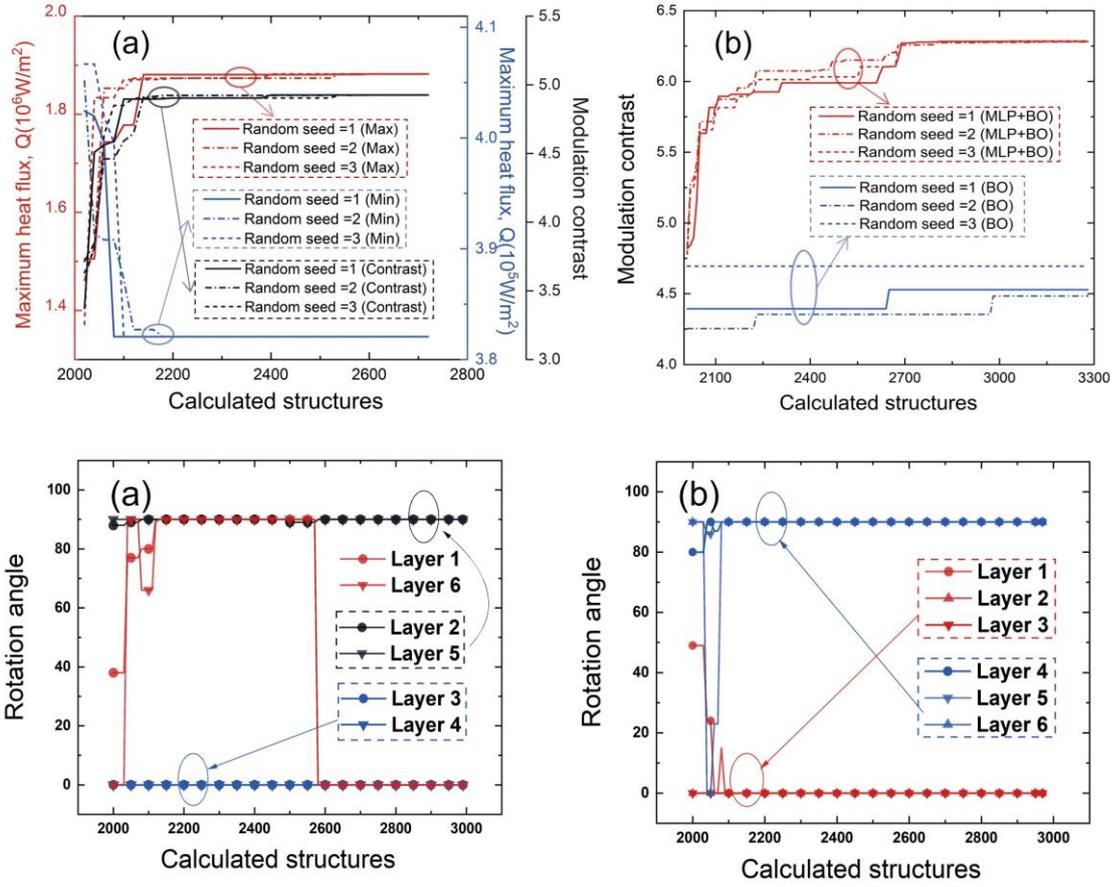

**Fig. 3.** (a)-(b) Modulation contrast varying with the number of calculated structures needed to find the optimal structures in the proposed optimization process: (a) step 1, (b) step 2. (c)-(d) show the rotation angle of each layer varying with the number of calculated structures, (c) maximum heat flux in step 1, (d) minimum heat flux in step 1.

**Table. 2.** Consumed time comparison of the proposed workflow for the design of multilayer metamaterials.

| Process | Accurate computation | Surrogate model training | optimization |
| --- | --- | --- | --- |
| Computation time | ~50 h / 3000 points | ~100 min | ~50 min |

To demonstrate the physical mechanism behind the larger modulation contrast caused by rotation, the spectral heat flux of the optimal structure is calculated and shown in Fig. 4. The black line refers to spectral flux after step 1 angular optimization, where the detail rotation angles are shown in Table 3. It can be observed that a significantly higher heat flux was achieved by α-MoO$_3$ when the vacuum gap distance $d$ was in the sub-micron range. Moreover, the heat flux was significantly enhanced in all three Reststrahlen bands regions when the rotation angles are [0 90 0 90 0 90], and a sharp peak also appears at $\omega$ = 1.8228×10$^{14}$ rad/s. When the rotation angles $\varphi_i$ are changed to [0 0 0 90 90 90], the heat flux within the Reststrahlen bands decrease significantly compared to that of former rotation angles. Specifically, when the rotation angles $\varphi_i$ are [0 0 0 90 90 90], a sharp decrease in heat flux was also observed in the spectral region of Reststrahlen band I and II. The enhancement of Reststrahlen bands and sharp peak heat flux are attributed to the excitation of HPPs and HSPhPs. In the following sections, the properties of HPPs and HSPhPs under different rotation angle, layer thickness and gap distance will be further discussed by examining the energy transmission coefficient.

**Table. 3.** The optimal design parameters of the multilayer hyperbolic metamaterials for near-field thermal radiative modulator with the highest modulation contrast ratio.

| Step 1 | $\varphi_1$ | $\varphi_2$ | $\varphi_3$ | $\varphi_4$ | $\varphi_5$ | $\varphi_6$ | heat flux/ kwm$^{-2}$ | modulation contrast |
|---|---|---|---|---|---|---|---|---|
| | 0 | 90 | 0 | 90 | 0 | 90 | 1882 | 4.93 |
| | 0 | 0 | 0 | 90 | 90 | 90 | 382 | |
| Step 2 | $t_1$/ nm | $t_2$/ nm | $t_3$/ nm | $t_4$/ nm | $t_5$/ nm | $t_6$/ nm | distance $d$ /nm | modulation contrast |
| | 1 | 65 | 11 | 11 | 62 | 4 | 100 | 6.29 |

Figures 5(a) and 5(b) show the energy transmission coefficient between the emitter and receiver as a function of wave vector component $k_x$ and $k_y$ at $\omega$ = 1.06×10$^{14}$ rad/s, and the corresponding rotation angles are [0 90 0 90 0 90] and [0 0 0 90 90 90], respectively. Here, $k_x$ and $k_y$ represent the projections of $\beta$ on the $x$-axis and $y$-axis, respectively. As indicated by bright colors in Figure 5(a), there is a large amount of heat flux near the center $k$ vector region which is associated with the excitation of HPPs. In fact, HPPs are volume modes that exhibit hyperbolic dispersion relations in anisotropic materials. The hyperbolic dispersion relation ensures that the mode has an infinite wave vector component in the hyperbolic materials. Therefore, it has been widely demonstrated that the significantly enhanced non-contact radiative heat transfer can be achieved between the hyperbolic uniaxial materials due to the excitation of HPPs. The conditions for hyperbolic phonon polarization in α-MoO₃ can be expressed as,

$$\frac{(\varepsilon_y k_y^2 + \varepsilon_x k_x^2)}{\varepsilon_z} < 0 \tag{4}$$

Assuming $k_y \gg k_0$, $k_x \gg k_0$, more specifically, when $\varepsilon_z > 0$, $\varepsilon_x < 0$, and $\varepsilon_y > 0$, the existence interval of HPPs can be expressed as,

$$-\sqrt{-\frac{\varepsilon_x}{\varepsilon_y}} < \frac{k_y}{k_x} < \sqrt{-\frac{\varepsilon_x}{\varepsilon_y}} \tag{5}$$

When $\varepsilon_z > 0$, $\varepsilon_x > 0$, and $\varepsilon_y < 0$, the HPPs can occur in the region,

$$\frac{k_y}{k_x} > \sqrt{-\frac{\varepsilon_x}{\varepsilon_y}} \quad \text{and} \quad \frac{k_y}{k_x} < \sqrt{-\frac{\varepsilon_x}{\varepsilon_y}} \tag{6}$$

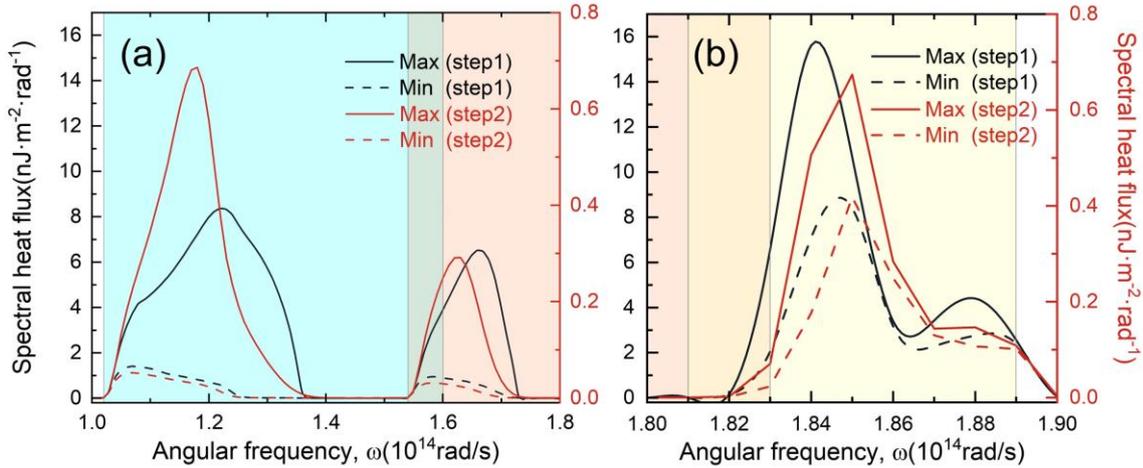

**Fig. 4.** Modulated maximum and minimum spectral heat flux as a function of angular frequency after the two-step optimization. (a) Reststrahlen band I, II region, and (b) band III region.

According to Eq. (4-6), at the angular frequency $\omega = 1.06 \times 10^{14}$ rad/s, when rotation angle $\varphi = 0$, the relative dielectric constants are $\varepsilon_x = -109 + j\,13.4$, $\varepsilon_y = 7.06 + j\,0.0193$, $\varepsilon_z = 2.76 + j\,0.000673$. By ignoring the imaginary part of $\varepsilon_m$ ($m = x, y, z$), the boundary curve of HPPs region in the $k_x$-$k_y$ plane can be obtained as $k_y = \pm 3.93 k_x$, and it becomes $k_y = \pm 0.254\, k_x$ when rotates by 90 degrees. The asymptotes for HPPs region are expressed by the purple dashed lines in Fig. 5(a) and (b). In Fig.5(a), When the rotation angle is [0 90 0 90 0 90], the HPPs excited by each layer occur in all directional angles, so that the bright color covers large area, resulting in large heat flux. The region bounded by $k_y = \pm 3.93k_x$ exhibits larger heat flux, which is predominantly influenced by the emitter layer that is located closest to the vacuum gap. Figure 5(b) shows that these asymptotes enclose the regions with high heat flux, as indicated by the bright colors. When the rotation angle is [0 0 0 90 90 90], the HPPs regions of the emitter and receiver partially overlap. In this case, large energy transmission coefficients only exist in the four overlapping regions indicated by the bright colors as shown in Fig. 5(b). Therefore, the heat flux is significantly reduced compared to the case shown in Fig 5(a). In other words, the increase comes from the interaction and combination of HPPs excited from different layers, and the decrease comes from the misalignment of HPPs between the emitter and receiver.

Figure 5(c) and (d) show the energy transmission coefficient between the emitter and receiver varying with the wave vector components $k_x$ and $k_y$ at $\omega = 1.2 \times 10^{14}$ rad/s. At $\omega = 1.2 \times 10^{14}$ rad/s, when rotation angle $\varphi=0$, the main values of the relative permittivity components are $\varepsilon_x = -5.3 + j\, 0.482$, $\varepsilon_y = 8.08 + j\, 0.0389$, and $\varepsilon_z = 2.82 + j\, 0.00105$. Therefore, on the $k_x$-$k_y$ plane, the boundary curve of HPPs region can be obtained as $k_y = \pm 1.38 k_x$, which becomes $k_y = \pm 0.72 k_x$ when rotates by 90 degrees. The asymptotes for HPPs region are expressed by the purple dashed lines in Fig. 5(c) and (d). In Fig. 5(c), the bright color clearly indicates the excited HPPs covers large area at all directional angles. The mechanism for large modulation contrast is similar to the case shown in Fig 5. (a) and (b). The modulation performance is more drastic in this area because of smaller overlap region in Fig 5. (b). Figure 5(e) and (f) show the energy transmission coefficient between the emitter and the receiver varying with the wavevector components $k_x$ and $k_y$ at $\omega =1.82 \times 10^{14}$ rad/s, at which a peak of heat flux appears. At this angular frequency, when rotation angle $\varphi = 0$, the permittivity components are $\varepsilon_x = 1.7 + j\, 0.0213$，$\varepsilon_y = -0.192 + j\, 0.0623$，$\varepsilon_z = -12.3 + j\, 1.96$. The dashed line can be described as $k_y = \pm 2.98\, k_x$ and $k_y = \pm 0.34 k$.

When $\varepsilon_z < 0$, $\varepsilon_x > 0$, and $\varepsilon_y < 0$, the HPPs can occur in the region,

$$-\sqrt{-\frac{\varepsilon_x}{\varepsilon_y}} < \frac{k_y}{k_x} < \sqrt{-\frac{\varepsilon_x}{\varepsilon_y}} \tag{7}$$

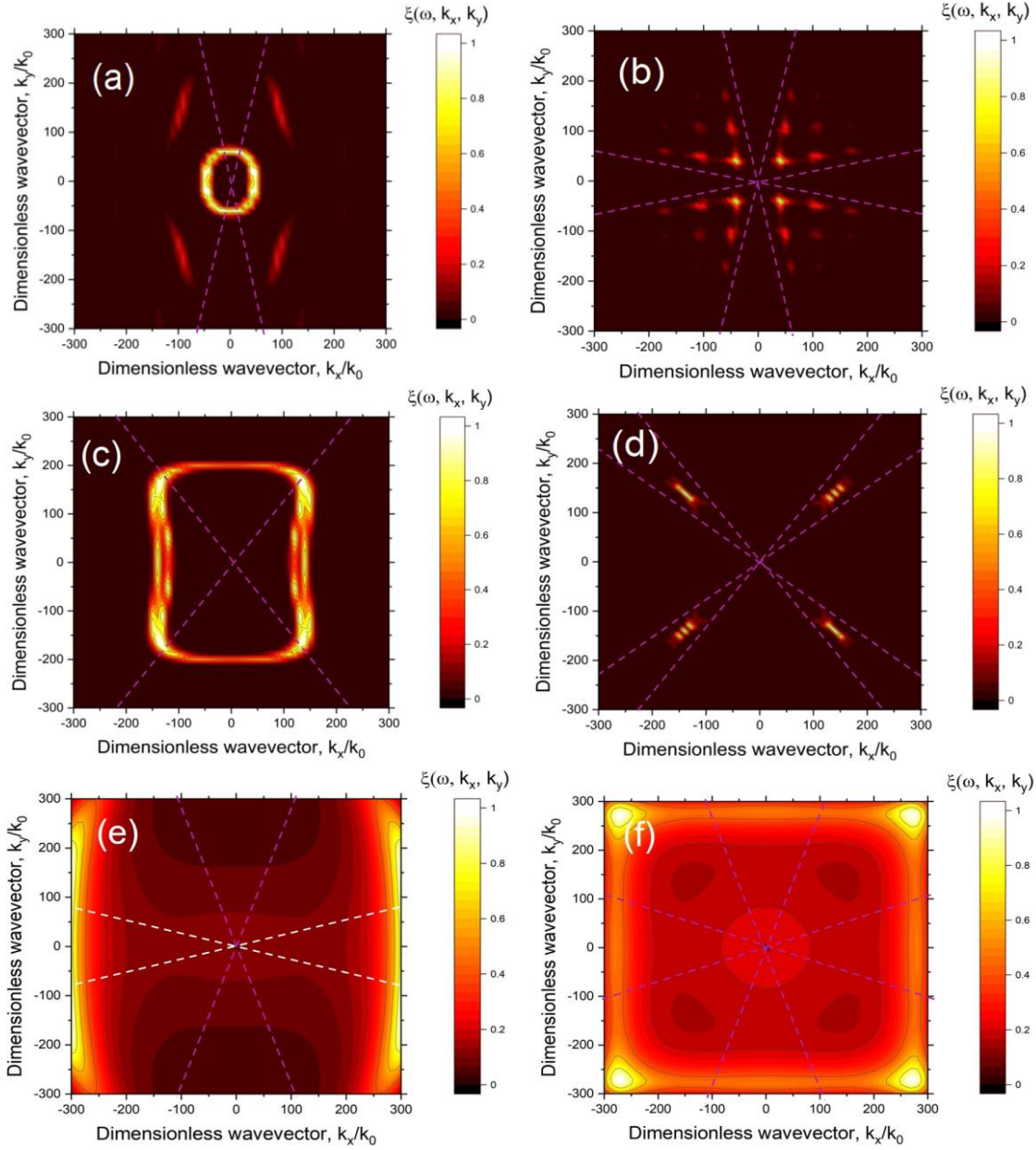

**Fig. 5.** The energy transmission coefficient contour plot as a function of wavevector components $k_x$ and $k_y$ for thermal modulator with maximum and minimum heat flux. The gap distance $d$ =20 nm, and the layer thickness $t$ =7 nm. (a) modulator with maximum heat flux at $1.06\times10^{14}$ rad/s, (b) modulator with minimum heat flux at $1.06\times10^{14}$ rad/s, (c) modulator with maximum heat flux at $1.2\times10^{14}$ rad/s, (d) modulator with minimum heat flux at $1.2\times10^{14}$ rad/s, (e) modulator with maximum heat flux at $1.82\times10^{14}$ rad/s, (f) modulator with minimum heat flux at $1.82\times10^{14}$ rad/s.

Based on the discussion above, the bright color in the region bounded by purple dashed line is attributed to HPPs. However, the high heat flux represented by the brighter colors inside the white line cannot be attributed to HPPs. Here, this enhancement of heat flux is caused by the resonance of the Dyakonov waves, a special type of surface wave that is excited at the interface between isotropic and anisotropic media. Due to the features of a hyperbolic surface in dispersion relation, the Dyakonov waves are called HSPhPs. For uniaxial and biaxial materials with the main relative permittivity component being positive, the Dyakonov waves have been extensively studied [44], which is the main contribution to the peak of heat flux. In fact, the dispersion curves have asymptotes in the form,

$$\frac{k_y}{k_x} = \pm\sqrt{-\frac{\varepsilon_x\varepsilon_z-\varepsilon_d^2}{\varepsilon_y\varepsilon_z-\varepsilon_d^2}}. \qquad (8)$$

When the rotation angle is [0 0 0 90 90 90], the enhanced heat flux is distributed in the region shown in Fig. 5(f). Due to the misalignment of HPPs and HSPhPs between the emitter and receiver, the heat flux is similar but smaller compared to the case shown in Fig. 5(e).

For the step 2 in the optimization process, the rotation angle is fixed at [0 90 0 90 0 90] and [0 0 0 90 90 90], the modulation contrast further reaches 6.29 by optimizing the thickness of each layer and the gap distance between emitter and receiver. The optimal parameters for step 2 are also listed in Table 3. The red line in Fig. 4 represents the spectral heat flux for optimal thermal modulator obtained in step 2. The modulation contrast increases significantly especially in Reststrahlen Band I region after the thickness and gap distance optimization.

Figure 6 shows the energy transmission coefficient counter plot varying with the wavevector components $k_x$ and $k_y$ at $\omega = 1.2\times 10^{14}$ rad/s. In Fig. 6 (a) and (b), the gap distance between emitter and receiver is 20 nm. By comparing Fig. 6(a) with Fig. 5(c), larger bright area occurs in the upper and lower area bounded by $k_y = \pm 0.72k_x$. The reason is that with the increase of layer thickness $t_2$, more HPPs are excited from the second layer of the emitter corresponding to the region bounded by $k_y = \pm 0.72k_x$. The combination of HPPs excited by different layers can largely enhance the heat flux compare with the single layer structure. For Fig. 6 (c) and (d), when the gap distance is 100 nm, there are smaller bright area since the NFRHT will become weaken for larger gap distance.

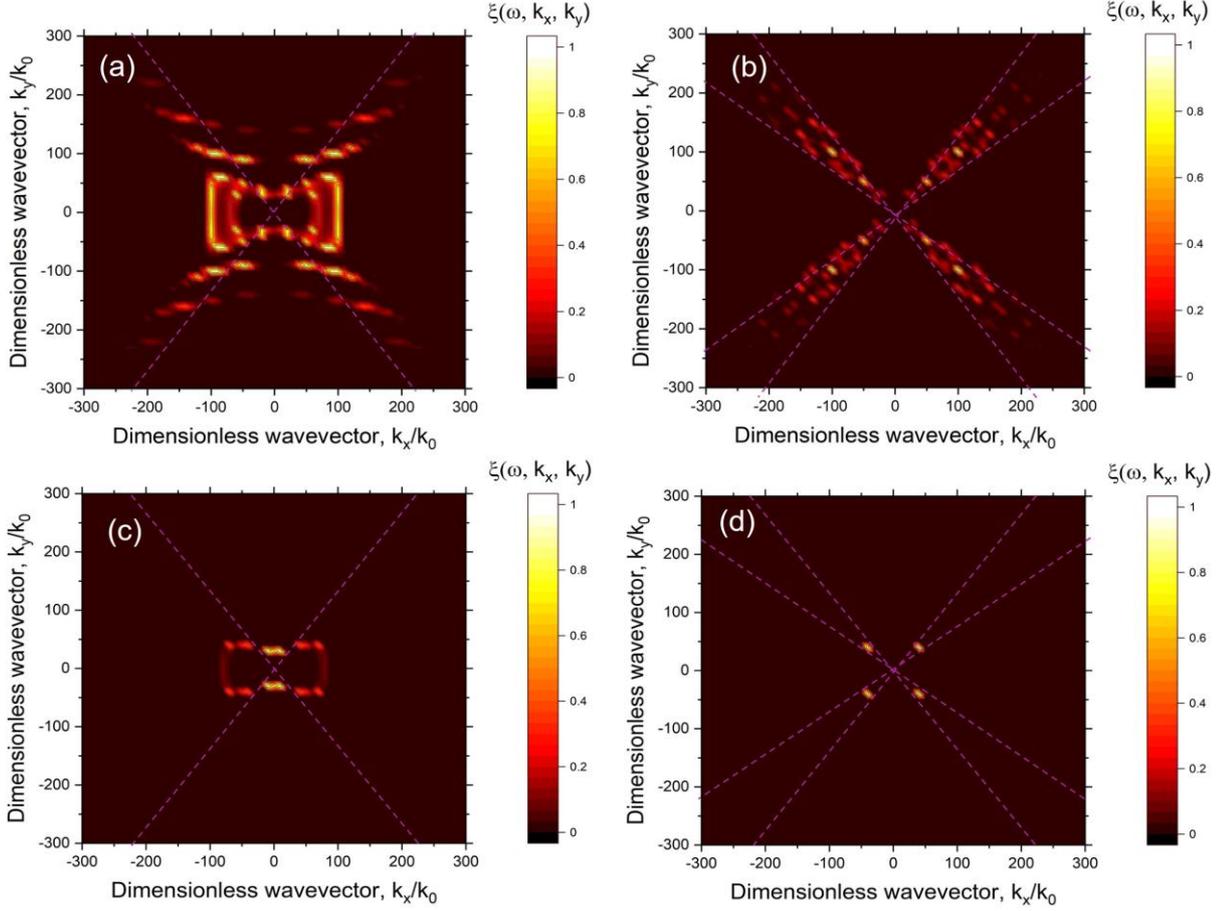

**Fig. 6.** The energy transmission coefficient counter plot of thermal modulator as a function of wavevector components $k_x$ and $k_y$ for modulator with maximum and minimum heat flux at $1.2 \times 10^{14}$ rad/s. The detail rotation angle and thickness are listed in step 2 of Table. 3. Modulator with maximum heat flux, (a) $d$=20 nm, (c) $d$=100 nm, modulator with minimum heat flux, (b) $d$=20 nm, (d) $d$=100 nm.

## 4. Conclusions

In conclusion, we have proposed the optimization design of a multilayer thermal modulator based on hyperbolic material α-MoO$_3$ using a hybrid machine learning workflow combining the MLP neural network and BO algorithm, to obtain the maximum rotation thermal modulation contrast ratio. The MLP neural network is trained to simulate the NFRHT heat flux/ modulation contrast between emitter and receiver, and the BO is used to find the maximum modulation contrast ratio. The thermal modulator composed of multilayer metamaterials by tuning mainly lattice rotation, layer thickness, and gap distance are systematically explored and discussed. The results show that after the optimization of rotation angles in step 1, the modulation contrast reaches 4.93, which is 54% higher than that of the single-layer modulator structure. With further optimization of the layer thickness and the gap distance in step 2, the modulation contrast ratio reaches 6.29, which is 97% higher than that of the single-layer structure. The behind physical mechanism we can learn from machine learning is that when the HPPs excited by different layers covers all directional angles, the NRFHT between emitter and receiver can be significantly enhanced by reasonable tuning of the rotation angle for each layer. With the HPPs and HSPhPs between the emitter is misaligned after rotation, a rather small modulated NRFHT heat flux can also be achieved, thereby a large modulation contrast is finally obtained. The findings of this research provide valuable guidance for the data-driven design of near-field radiation devices considering large number degrees of freedom.

**Declaration of Competing Interest**

The authors declare that they have no known competing financial interests or personal relationships that could have appeared to influence the work reported in this paper.

**Acknowledgements**

This work was supported by the Shanghai Pujiang Program (No. 20PJ1407500), National Natural Science Foundation of China (No. 52006134), Shanghai Key Fundamental Research Grant (No. 21JC1403300). The computations in this paper were run on the Siyuan-1 cluster supported by the Center for High Performance Computing at Shanghai Jiao Tong University.